%% file: main.tex
\documentclass{article}

\usepackage{PRIMEarxiv}

\usepackage{amsmath}
\usepackage[utf8]{inputenc} 
\usepackage[T1]{fontenc}    
\usepackage{hyperref}       
\usepackage{url}            
\usepackage{booktabs}       
\usepackage{amsfonts}       
\usepackage{nicefrac}       
\usepackage{microtype}      
\usepackage{lipsum}
\usepackage{fancyhdr}       
\usepackage{graphicx}       
\graphicspath{{images/}}     

\title{Separable spatio-temporal kriging \\ for fast virtual sensing
}

\author{
  Michele Lambardi di San Miniato, Ruggero Bellio, Luca Grassetti, Paolo Vidoni \\
  Department of Economics and Statistics \\
  University of Udine \\
  Udine, Italy \\
  \texttt{\{michele.lambardi, ruggero.bellio, luca.grassetti, paolo.vidoni\}@uniud.it} 
}

\usepackage{siunitx} 

\usepackage{subcaption} 
\renewcommand{\vec}{\mathrm{vec}} 

\begin{document}
\maketitle

\begin{abstract}
Environmental monitoring is a task that requires to surrogate system-wide information with limited sensor readings. Under the proximity principle, an environmental monitoring system can be based on the virtual sensing logic and then rely on distance-based prediction methods, such as $k$-nearest-neighbors, inverse distance weighted regression and spatio-temporal kriging. The last one is cumbersome with large datasets, but we show that a suitable separability assumption reduces its computational cost to an extent broader than considered insofar. Only spatial interpolation needs to be performed in a centralized way, while forecasting can be delegated to each sensor. This simplification is mostly related to the fact that two separate models are involved, one in time and one in the space domain. Any of the two models can be replaced without re-estimating the other under a composite likelihood approach. Moreover, the use of convenient spatial and temporal models eases up computation. We show that this perspective on kriging allows to perform virtual sensing even in the case of tall datasets.
\end{abstract}

\keywords{Spatio-temporal kriging \and Distance-based prediction \and Separability \and Isotropy \and Indoor environments \and Composite likelihood \and Distributed calculus}

\section{Introduction}

Environmental monitoring systems rely on virtual sensing logic to predict relevant variables of their target environment. While the information on the whole system is of interest, this is typically based on sensor readings, which are limited in both space and time, so it is necessary to surrogate them, based on some suitable statistical method \cite{Gramacy2020}. Variables of interest may include, for instance, room temperature \cite{Nguyen2017}, energy consumption \cite{Carpenter_2018} and air quality \cite{Liu_2018}. We consider the case of enclosed environments \cite{Li2011,Nguyen2017,Liu_2018}, as contrasted to other applications that are aimed at larger environments like ecosystems \cite{Rodriguez_1974,Mardia1993}. Moreover, our focus is on applications that need real-time control \cite{Dong2013}.

\begin{figure}
	\centering
	\includegraphics[width=0.4\linewidth]{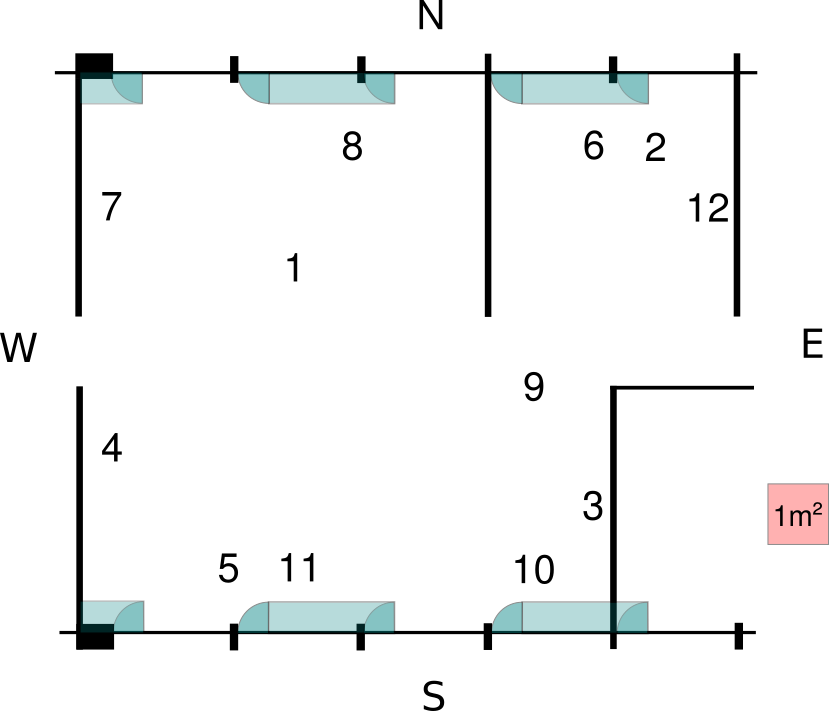}
	\caption{\label{fig:room} Room with windows, cardinal points and sensor locations, modified from the original map \cite{Brunello2021}.}
\end{figure}

The motivating example for this work comes from a virtual sensing project at Silicon Austria Labs GmbH, a European research center for electronic based systems \cite{Brunello2021}. The data relate to an office room in Villach, Austria, that has been monitored for 19 weeks between October 2019 and March 2020. The temperature (in \si{\degreeCelsius}) is reported by twelve sensors every 10 seconds, along with other 
physical measurements like pressure 
(in \si{\pascal}) and light (in \si{\lux}). The room is \SI{127}{\metre\squared} large and it is structured as reported in Figure \ref{fig:room}. The picture also shows the locations of sensors and windows, along with the cardinal points.

The twelve sensors are all Raspberry Pi Zero boards. Their measurements are broadcast over a wireless network to a database server, which is a Raspberry Pi 3 instead. Raspberry Pis are popular and affordable single-board computers that are widely used in home automation, smart systems \cite{Ferdoush_2014}. This example has some key aspects, including data referenced both temporally and spatially, high-frequency measurements, resulting in multivariate times series with as many as $10^6$ observations for each sensor. The server has a limited yet non-negligible computational power, which is also important, as it allows to process data locally if this task is planned carefully, by taking into consideration the limitations of the monitoring system.

As common in modern data analysis, there are at least two main and opposite approaches to deal with sensor data. These two opposites are represented by interpretable models and black-box algorithms, respectively. The former include specifications based on actual knowledge about physical aspects of the system \cite{Dong2013}, often hard to formulate; the latter include neural networks and other machine learning techniques that achieve remarkable performance levels and are readily available in general software. Other authors addressed the same datasets of our present analysis \cite{Brunello2021}, but they used techniques such as XGBoost regression \cite{Chen_2016} and LSTM recurrent neural networks \cite{Hochreiter_1997}. These methods do not provide interpolation by design, but they can do so only after some suitable engineering, thus some generalization issues emerge.

Here we advocate for an approach lying between the two extremes, which is statistically sound without compromising prediction performance. We are interested in simple models and distributed computing, thus on scalable methods that leverage on the proximity principle. It is then natural to resort to distance-based prediction methods \cite{Oktavia2016}, which include for instance inverse distance weighted regression (IDWR), $k$-nearest-neighbors ($k$-NN) \cite{Azzalini2012} and spatio-temporal kriging \cite{Aryaputera_2015}. These methods are somewhat related to pure spatial data analysis \cite{Bivand_2013}. Our focus is on kriging, in particular. This approach relies on a correlation model, which depends on distances between measurements in time and space. A crucial assumption for distributed calculus is spatio-temporal separability, which implies two separate models for spatial and temporal correlations \cite{Mardia1993}. This assumption is hardly suitable for large environments, where some locations can anticipate events that will occur somewhere else. In smaller environments, separability provides instead a useful approximation that catches up with more complicated, non-separable models \cite{Genton_2007}.

While involving a simple model, kriging is cursed by the enormous cost of computation, mostly due to the inversion of large matrices. Some approximations have been devised to make kriging tractable like, for instance, covariance tapering \cite{Kaufman}. A composite likelihood approach can be used to estimate a separable model, which allows to separate the estimation of the spatial model from the estimation of the temporal model. Also, some models in the time domain can spare the cumbersome matrix inversions and thus simplify both estimation and prediction. For instance, auto-regressive models can be estimated just by minimizing the conditional sum of squares, and they come with compact forecasting rules \cite{Box}. As to spatio-temporal predictions, we show that these can be seen as a spatial interpolation of temporal forecasts under separability, which allows to leverage on specific advantages of time series and spatial models. All these possibilities seem somewhat overlooked in modeling-aware literature, despite the attention received by separability itself.

The plan of the paper is as follows. In Section 2, we recall basic kriging formulation, while emphasizing some correlation structures of practical importance. In Section 3, we detail our inferential and predictive framework, focusing on distributed computing. Section 4 illustrates the application of the proposed methodology to the motivating example, whereas Section 5 is devoted to some possible twists and extensions. Finally, Section 6 presents some concluding remarks.

\section{Model specification}

We deal with the case of data that are both spatially and temporally referenced, so we use a space index $s$ and a time index $t$. The space index $s$ takes its value in $\mathcal{S}=\{1,\dots,S\}$ and is a pointer to one out of $S$ locations in space, while the time index $t$ takes its value in $\mathcal{T}=\{1,\dots,T\}$ and is a pointer to one out of $T$ time frames. A joint index $st$ denotes location $s$ at time $t$. Since dealing with a constant sampling rate, we consider a discrete time system with equispaced time frames. In the long run, it holds $T \gg S$.

Let $d_{s,s'}$ be a spatial distance, defined for all pairs of locations $s,s'\in\mathcal{S}$, thus endowed with non-negativity, symmetry and triangle inequality. Distance is ordinarily evaluated along straight lines in the absence of physical obstacles; otherwise, the length of the shortest path is considered. We choose the Euclidean distance for this purpose. Temporal distance is defined analogously \cite{Franceschetti_2007} as
$$d_{t,t'} = |t - t'| \,.$$
Here, $d_{s,s'}$ and $d_{t,t'}$ are the generic elements of the $S\times S$ spatial distance matrix $d_S$ and of the $T\times T$ temporal distance matrix $d_T$, respectively.

The data $y$ are structured as follows:
$$y=\begin{bmatrix}
	y_{11} & \dots & y_{S1} \\
	\vdots & y_{st} & \vdots \\
	y_{1T} & \dots & y_{ST}
\end{bmatrix} \,,$$
so the data related to the location $s$ are all stored in the same column, while those related to the time frame $t$ are all stored in the same row. As new data are observed, they are appended to $y$ as a new row. The data $y$ are modeled by the random matrix $Y$ and the mean matrix $\mu$ with the same number of rows and columns of $y$.

Let $\vec$ be a unary operator defined for matrices that stacks their columns into a single vector \cite{Hartwig_1975}. $Y$ is assumed to be a multivariate normal with scale parameter $\sigma$ and correlation matrix $R$, in the sense that $R$ is the correlation matrix of $\vec(Y)$. More formally, we assume that $Y$ has the following density function.
\begin{equation}\label{eq:fulllik}
	f(y;\mu,\sigma,R) = \frac{1}{\sqrt{\det(2\pi\sigma^2 R)}} \cdot \exp\left\{-\frac{1}{2}\vec\left(\frac{y-\mu}{\sigma}\right)^\top R^{-1}  \, \vec\left(\frac{y-\mu}{\sigma}\right)\right\} \,.
\end{equation}

We assume that $\mu_{st}$ is a smooth function of $t$ and shared across locations, so it makes sense to estimate it with an asymmetric moving average $m_t$, defined below, which pools data across locations from $w$ time frames preceding $t$. Namely,
\begin{equation}\label{eq:movavg}
	\hat{\mu}_{st} = m_t \,,\quad \text{with} \quad m_t = \frac{1}{S \cdot w}\sum_{s=1}^S \sum_{i=1}^w Y_{s(t-i)} \,.
\end{equation}
We set $w$ equal to the number of observations per sensor in the 24 hours. Thus, the latest estimate available for $\mu$ can also serve as an estimate for future trend $\mu'$, assuming stability in the short term. Such an assumption can be credible in cases where the univariate time series may agree on a single latent factor ruling all of them.

The parameter $\sigma>0$ contributes only in making predictions probabilistically calibrated \cite{Berrocal_2007}, because it is just a scale parameter, like the error standard deviation in classical linear regression, thus involved in prediction variance but not in mean predictions; see Appendix \ref{app:krig}. A simple estimator of $\sigma$ is the following, based on the assumption of constant variance through time and space.
$$\hat{\sigma}=\sqrt{\frac{1}{ST}\sum_{s=1}^S \sum_{t=1}^T (Y_{st} - \hat{\mu}_{st})^2} \,.$$

We resort to the classical kriging approach, which belongs to frequentist statistics, but this methodology also has a Bayesian counterpart involving a prior distribution on parameters.\cite[Ch. 5--7]{Diggle_2007} As per kriging approach, we assume that correlations between components of $Y$ are stationary and thus depend on their distances in space and time. The covariance between any two components of $Y$, say, $Y_{st}$ and $Y_{s't'}$, is modeled as
$$\mathrm{cov}(Y_{st},Y_{s't'}) = \sigma^2 \cdot \mathrm{cor}_S(d_{s,s'}) \cdot \mathrm{cor}_T(d_{t,t'}) \,,$$
where $\mathrm{cor}_S(\cdot)$ is the spatial auto-correlation function (ACF) \cite{Cressie_1993}, while $\mathrm{cor}_T(\cdot)$ is the temporal ACF \cite{Box}. The product between spatial and temporal correlations is implied by the separability assumption. ACFs depending on distances and not directions are implied by an isotropy assumption. Both separability and isotropy can simplify modeling and computing \cite{Rodriguez_1974,Mardia1993,Zhang2015,Nguyen2017,Gramacy2020}.

For the sake of illustration, the spatial ACF can be, for instance, one of the following \cite{Franceschetti_2007,Bivand_2013,Oktavia2016}:
\begin{itemize}
	\item{Matérn ACF \cite{Gaetan_2010}}
	$$\mathrm{cor}_S(d) = \frac{2^{1-\alpha}}{\Gamma(\alpha)} (d/\lambda)^\alpha \mathrm{K}_\alpha(d/\lambda) \,,\quad \lambda,\alpha>0 \,,$$
	with $\Gamma(\cdot)$ the gamma function and $\mathrm{K}_\alpha(\cdot)$ the modified Bessel function of the second kind;
	\item{power exponential ACF \cite{Gramacy2020}}
	$$\mathrm{cor}_S(d) = \exp\{-(d/\lambda)^\beta\} \,,\quad \lambda,\beta > 0 \,.$$
\end{itemize}
Here, $\lambda$ can be regarded as a \textit{range} parameter. We refer to $\alpha$ and $\beta$ as \textit{smoothness} parameters. Both Matérn and power exponential ACFs include two notable sub-cases:
\begin{itemize}
	\item when $\alpha=1/2$ and $\beta=1$, the exponential ACF is implied \cite{Rodriguez_1974};
	\item when $\alpha\to\infty$ and $\beta=2$, the Gaussian ACF is implied \cite{Gramacy2020,Maes_2021}, which is also known as squared-exponential ACF \cite{Stachniss_2009,Liu_2018,Carpenter_2018} and involved in the double-stable model \cite{Nguyen2017}.
\end{itemize}

The temporal ACF in discrete time can be, for instance, one of the following \cite{Box}:
\begin{itemize}
	\item{ACF of a stationary auto-regressive model of order 1} $$\mathrm{cor}_T(d) = \phi^{|d|} \,,\quad |\phi|<1\,;$$
	\item{ACF of an invertible moving-average model of order 1} $$\mathrm{cor}_T(d) = \left\{\begin{array}{lr}
		1 & d=0 \\
		\alpha & d=\pm 1 \\
		0 & \text{otherwise}
	\end{array}\right. \,,\quad |\alpha|<1/2 \,.$$
\end{itemize}
More complicated ACFs are possible when looking at more flexible time series models, such as the multiplicative seasonal AR models that are used in the empirical application presented later. Some approaches assume weak or no correlation structure, like empirical kriging \cite{Aryaputera_2015}. These approaches are necessarily less scalable but may still work for suitably targeted tasks.

Let $\mathrm{cor}_S(\cdot)$ and $\mathrm{cor}_T(\cdot)$ be vectorized function, that is, they transform matrices in an entry-wise fashion. Then, $R_S=\mathrm{cor}_S(d_S)$ will be a spatial correlation matrix and $R_T=\mathrm{cor}_T(d_T)$ a temporal correlation matrix. We call
\begin{equation}\label{eq:kron}
	R = R_S \otimes R_T \,,
\end{equation}
the spatio-temporal correlation matrix. Here, $\otimes$ denotes the Kronecker product.

Both temporal and spatial ACFs can be modified in order to account for noisy data by including a so-called nugget effect \cite{Gaetan_2010,Gramacy2020}. This means that the spatial ACF $\mathrm{cor}_S(d_S)$ and the temporal ACF $\mathrm{cor}_T(d_T)$ are multiplied by $\beta_S$ and $\beta_T$ when $d_S=0$ and $d_T=0$, respectively, the parameters $\beta_S$ and $\beta_T$ taking values in the interval $]0,1]$ \cite{Aryaputera_2015}. We refer to $1-\beta_S$ and $1-\beta_T$ as the \textit{nugget} parameters, in space and time domains, respectively. Some authors apply the nugget directly to the spatio-temporal covariance function, but this will break up separability \cite{Banerjee_2008,Zhang2015}. The latter way of modeling is more natural in the case of additive covariance models \cite{Dumelle_2021}.

\section{Inferential aspects}

This section presents two strategies that allow to perform estimation and prediction under a separable model, with a low computational cost. In particular, we base estimation on a novel composite likelihood approach. Then, leveraging on the peculiar expression of the kriging mean formula, we show how to compute predictions efficiently under separability.

\subsection{Estimation}

The distribution of $Y$ in Equation \eqref{eq:fulllik} is assumed to be indexed by a parameter vector $\theta$ via $\mu,\sigma,R$. Let $\theta$ be partitioned as $\theta=(\mu^\top,\sigma,\psi^\top)^\top$, with $\psi$ the correlation parameters ruling $R$. Moreover, $\psi$ can be partitioned as $\psi=(\psi_S^\top,\psi_T^\top)^\top$, where $\psi_S$ and $\psi_T$ are the spatial and temporal correlation parameters, respectively. In particular, $R_S$ depends only on $\psi_S$, while $R_T$ depends only on $\psi_T$. As a starting point, we consider the likelihood function $\mathcal{L}(\theta;y)$, defined as $$\mathcal{L}(\theta;y) = f(y;\mu,\sigma,R)\,.$$ Let the sample correlation matrix be defined as rank-$1$ matrix
$$\hat{M}= \frac{1}{\hat{\sigma}^2}\vec(y-\hat{\mu}) \vec(y-\hat{\mu})^\top \,,$$
where $\mu$ and $\sigma$ have been replaced with some estimates, denoted by $\hat{\mu}$ and $\hat{\sigma}$, respectively. When replacing $\mu$ and $\sigma$ with estimates, the likelihood function turns into a so-called pseudo likelihood $\mathcal{L}^p(\psi;y)$, which allows to make inference on $\psi$ alone \cite{Gong_1981,Gaetan_2010}, defined as
\begin{equation}\label{eq:pseudo1}
	\mathcal{L}^p(\psi;y) = |R|^{-1/2} \exp\left\{-\frac{1}{2} \mathrm{trace}\left(R^{-1} \hat{M}\right)\right\} \,.
\end{equation}

Kriging is often cumbersome due to the inversion of $R$, and actually the pseudo likelihood in Equation \eqref{eq:pseudo1} is intractable with high-dimensional data. Separability reduces the dimensionality of the problem, as it holds
$$R^{-1} = (R_S \otimes R_T)^{-1} = R_S^{-1} \otimes R_T^{-1} \,,$$
so two smaller inverse matrices must be computed instead of a single and larger one. However, as $T \gg S$, inverting $R_T$ alone can also be difficult.

Our proposal is two-fold. First, a marginal composite likelihood approach \cite{Caragea_2007} can be used, exploiting separability more in depth so that the tasks of estimating $\psi_S$ and $\psi_T$ can even be tackled with separately. Second, a suitable time-series model can help handling the temporal correlation implicitly, in the sense that $R_T$ needs not be evaluated at all, and make it possible to address tall data and high sampling rates.

Composite likelihoods are known in spatial statistics mainly as tools that simplify estimation and inference, like the pairwise composite likelihood \cite{Varin}, and they can also be used in model selection \cite{VarinVidoni}. Composite likelihoods allow to make inference on under-specified models but, even in the case of fully specified models, like in kriging, a suitable composite likelihood can reduce the computational cost of estimation. The estimator based on the full model can be computed in some cases, but it would be cumbersome with the dataset under investigation. We remark the tractability of our estimator by carrying out inference based on bootstrap. Cheaper computation comes at a price, since the estimator is naturally sub-optimal with respect to the maximum likelihood estimator. Nonetheless, the loss of efficiency might be not significant when dealing with high-frequency data.

With composite likelihoods, as with any so-called pseudo likelihood, estimation variance cannot be assessed as with classical likelihood functions, that is, based on the Hessian matrix. Parametric bootstrap can be used \cite{Davison_1997} in the case of kriging because the model is fully specified, and one can simulate datasets based on it. It is straightforward to simulate datasets under separability and some temporal models. We detail a bootstrap strategy in Appendix \ref{app:bootstrap}.

Within a single time frame $t\in\mathcal{T}$, it holds $R = R_S$, as per Equation \eqref{eq:kron}. Let the spatial composite likelihood be defined as
$$\mathcal{L}_S(\theta;y) = \prod_{t\in\mathcal{T}} f(y_{t};\mu_{t},\sigma,R_S) \,,$$ with $y_t$ and $\mu_t$ the data and mean vector at time frame $t$. The expression is the same of a ``small blocks" marginal composite likelihood \cite{Caragea_2007}. This composite likelihood, when replacing $\mu_t$ and $\sigma$ with estimates, turns into a pseudo likelihood for the spatial correlation parameters $\psi_S$, which is defined as
\begin{equation}\label{eq:spatpseudo}
	\mathcal{L}^p(\psi_S;y) = |R_S|^{-T/2} \exp\left\{-\frac{T}{2} \mathrm{trace}\left(R_S^{-1} \hat{M}_S\right)\right\} \,.
\end{equation}
Here, $\hat{M}_S$ is the sample correlation matrix between the univariate time series at distinct locations, defined as
\begin{equation}\label{eq:spatcor}
	\hat{M}_S= \frac{1}{T\hat{\sigma}^2}(Y-\hat{\mu})^\top (Y-\hat{\mu}) \,.
\end{equation}
The true correlations are better reflected by $\hat{M}_S$ if it is standardized to have a unit-valued diagonal, at least during the estimation step.

The estimator $\hat{\psi}_S$ for the spatial correlation parameters $\psi_S$ can be defined as the maximizer of the spatial pseudo likelihood $\mathcal{L}^p(\psi_S;y)$. This estimator is consistent, assuming $S$ is bounded and $T$ increases. If $R_S$ is a smooth function of $\psi_S$, the estimator can be seen as the solution of a system of estimating equations, the equation for the generic parameter $\alpha \in \psi_S$ being defined as follows:
$$\alpha \,:\, \mathrm{trace}\left\{\left(R_S - \hat{M}_S\right) \frac{\partial R_S^{-1}}{\partial \alpha} \right\} = 0 \,.$$
The above equation looks like a weighted average of the equations from the over-determined system $R_S - \hat{M}_S = 0$.

It is worth noticing the close resemblance between the pseudo likelihoods in Equations \eqref{eq:spatpseudo} and \eqref{eq:pseudo1}, but the correlation matrices involved are strikingly different, as $R_S$ is much smaller than $R$ in practical applications.

The temporal correlation parameter $\psi_T$ could be estimated analogously to $\psi_S$, by defining a temporal composite likelihood $\mathcal{L}_T$ as
\begin{equation}\label{eq:timelikelihood}
	\mathcal{L}_T(\theta;y) = \prod_{s\in\mathcal{S}} f(y_{s};\mu_{s},\sigma^2R_T) \,,
\end{equation}
with $y_s$ and $\mu_s$ the univariate time series and the mean vector of location $s$. The strategy is again to define a marginal composite likelihood \cite{Caragea_2007}. A temporal pseudo likelihood is obtained by replacing $\mu$ and $\sigma$ with estimates: this operation will completely disentangle the estimators of spatial and temporal parameters from each other. As contrasted to $R_S$, $R_T$ will be high-dimensional, so it is even more pressing the need to use a convenient correlation structure in the time domain. In particular, we resort to an auto-regressive model (AR) with multiple overlapping seasonal lags. For its definition, the classical lag operator $B$ can be introduced, such that
$$B Y_{st} = Y_{s(t-1)} \,,$$
and $B^\Delta Y_{st} = Y_{s(t-\Delta)}$ more in general. Then, letting $\epsilon_s$ be a white noise time series process, the multiplicative seasonal auto-regressive model is defined as follows \cite{Box}.
\begin{equation}\label{model:ar}
	\left[\prod_{k=1}^K (1-\phi_k B^{\Delta_k}) \right] \cdot (Y_s - \mu_s) = \epsilon_s \,,
\end{equation}
where $\Delta_1,\dots,\Delta_k,\dots,\Delta_K$ are the structural lags, and $\phi_1,\dots,\phi_k,\dots,\phi_K \in]-1,+1[$ the model coefficients.

Equation \eqref{model:ar} reduces the temporal correlation parameter to $\psi_T=(\phi_1,\dots,\phi_K)^\top$. AR modeling also makes it easy to approximately maximize the likelihood by minimizing the conditional sum of squares \cite{Box}, which can be attained via the coordinate descent method \cite{Wright_2015}. In this sense, we define
$$V_s^{-h} = \left[\prod_{k\neq h} (1-\phi_k B^{\Delta_k}) \right] \cdot (Y_s - \mu_s) \,,$$
and, by Equation \eqref{model:ar}, it holds
$$(1- \phi_h B^{\Delta_h}) V_s^{-h} = \epsilon_s \,,$$
so, the transformed time series $V_s^{-h}$ satisfies
$$\phi_h=\mathrm{cor}(V_{s}^{-h}, B^{\Delta_h}V_s^{-h})\,.$$
This relation motivates an iterative procedure that loops over the correlation parameters, and repeats until convergence. For each $\phi_h$, one evaluates the process $V_s^{-h}$ and then updates $\phi_h$ as the ACF of $V_s^{-h}$ at lag $\Delta_h$.

An online learning approach may be considered as an alternative. For instance, estimates can be updated continually via batch learning \cite{Azzalini2012} or exponentially-weighted moving means and covariances \cite{Hunter_1986,Huwang_2007}, which require additional tuning.

\subsection{Prediction}

Separability assumptions allow to simplify also prediction, as we show in this section. To this end, we must expand our notation. Then, let $Y'$ be from unobserved locations or times and with the mean matrix $\mu'$, while previously introduced matrices $Y$ and $\mu$ are now related to the available data $y$. The correlation matrix of $\vec(Y')$ is $R'$, whereas $R$ is the correlation matrix of $\vec(Y)$. The cross-correlation matrix between $\vec(Y')$ and $\vec(Y)$ is generally not square and it is denoted by $\rho$, which is based on a suitable joint distance matrix. Like $R$ in Equation \eqref{eq:kron}, also $R'$ and $\rho$ can be decomposed via Kronecker product as, respectively,
\begin{equation}\label{eq:otherkronecker}
	R'=R_S' \otimes R_T'\,,\quad \rho=\rho_S\otimes \rho_T\,.
\end{equation}
As typical in kriging, we treat $Y$ and $Y'$ as jointly normally distributed.

Kriging revolves around the conditional distribution of $Y'$ given $Y=y$ \cite{Gramacy2020}, though it was originally motivated as the linear unbiased prediction that is optimal with respect to the squared prediction error criterion \cite{Cressie_1990}. Conditional to $Y=y$, the distribution of $Y'$ is multivariate normal. The conditional mean matrix of $Y'$ and the conditional variance-covariance matrix of $\vec(Y')$, denoted by $\hat{y}'$, respectively, satisfy the following conditions.
\begin{equation}\label{eq:kriging}
	\vec(\hat{y}') = \vec(\mu') + \rho R^{-1} \vec(y - \mu) \,,\quad R'_{cond} = R' - \rho R^{-1} \rho^\top \,.
\end{equation}

Let $\beta_S$ and $\beta_T$ be spatial and temporal regression coefficients, defined as
\begin{equation}\label{eq:regcoef}
	\beta_S = \rho_S R_S^{-1} \,,\quad \beta_T = \rho_T R_T^{-1} \,,
\end{equation}
then the kriging mean formula in Equation \eqref{eq:kriging} can be written as
\begin{equation}\label{eq:krigspat}
	\hat{y}' - \mu' = \hat{z}\beta_S^\top \,,
\end{equation}
where $\hat{z}$ is a matrix of centered temporal forecasts only, defined as
\begin{equation}\label{eq:krigtemp}
	\hat{z} = \beta_T(y - \mu) \,.
\end{equation}
We prove this fact in Appendix \ref{app:krigmean}. The above result allows for at least two interesting uses.

First, our result allows for distributed calculus in kriging. Indeed, spatio-temporal prediction under separability can be carried out in two steps. The first step, in Equation \eqref{eq:krigtemp}, is temporal forecasting only within sensor locations. The second step, in Equation \eqref{eq:krigspat}, is the spatial interpolation of such forecasts for the needed locations. Then, spatio-temporal predictions can be seen as spatial interpolation of temporal forecasts. Geometrically, instead of moving along straight lines between spatio-temporal points, we cross one domain at a time. So, a separability assumption allows to separate domains also in prediction. Distributed calculus can be used for evaluating Equation \eqref{eq:krigtemp}, as each univariate time series is transformed separately by the matrix product. If the locations of interest are fewer than sensors, it will be more convenient to anticipate the interpolation step and thus to forecast only afterwards.

Correlation models affect prediction only through $\beta_S$ and $\beta_T$. So, any specification of a time model or a space model can be employed if it implies tractable prediction. This view motivates, for instance, the use of general interpolators, or state space models \cite{Hyndman_2002}, or integrated auto-regressive time series models \cite{Box}, that allow for simple prediction since $\beta_T$ in Equation \eqref{eq:regcoef} is sparse and explicit. The rather general auto-regressive moving-average model (ARMA) has already been considered by \cite{Ma_2005} though the MA component makes the model harder to estimate.

In applications, one may consider adding covariates to the analysis. Assume that there are $p$ variables indexed by $j=1,\dots,p$, so $Y_{jst}$ denotes the $j$-th variable at spatio-temporal coordinates $st$. In this cases, the marginal means $\mu_{jst}$ will likely depend on $j$. The correlation structure can be simplified according to a fully-factored model \cite{Mardia1993}, such that
$$\mathrm{cov}(Y_{jst}, Y_{j's't'}) = \gamma_{j,j'} \cdot \mathrm{cor}_S(d_{s,s'}) \cdot \mathrm{cor}_T(d_{t,t'}) \,.$$
Here, the new quantity $\gamma_{j,j'}$ represents the cross-sectional covariance between the $j$-th and $j'$-th variables at the same time and location. Thus, under a fully-factored model, our kriging mean formula scales easily, as a set of regression coefficients $\beta_C$ is defined besides $\beta_S$ and $\beta_T$, implied by the newly added domain of covariates.

We provided a simple expression for mean predictions, along with some optimization strategies. For the sake of completeness, we now illustrate how to compute prediction variances. In our view, the model can be used to design a prediction rule, resulting in mean prediction, while prediction variance is a performance measure, which can for instance be evaluated on a test set. This approach may be favored when the focus is especially on prediction.

Let $V$ be a matrix with the same size as $Y'$ with the entrywise variances of $Y'$ conditional to $Y=y$. Then, $\mathrm{vec}(V) = \sigma^2 \mathrm{diag}(R_{cond}')$, so it holds
\begin{equation}\label{eq:krigvar2}
V=\sigma^2\left\{\mathrm{diag}(R_T') \, \mathrm{diag}(R_S')^\top - \mathrm{diag}(R_T' - R_{T,cond}') \, \mathrm{diag}(R_S' - R_{S,cond}')^\top\right\} \,,
\end{equation}
where $\mathrm{diag}$ is the operator that returns the diagonal of a square matrix as a column vector, and $R_{S,cond}'$ and $R_{T,cond}'$ are defined as follows.
\begin{equation}\label{eq:condcors}
	R_{S,cond}' = R_S' - \rho_S R_S^{-1} \rho_S^\top \,,\quad R_{T,cond}' = R_T' - \rho_T R_T^{-1} \rho_T^\top \,.
\end{equation}
We prove this result in Appendix \ref{app:krigvar}. Equation \eqref{eq:krigvar2} has an advantage over \eqref{eq:kriging}, as the expression for $R_{T,cond}'$ is simple to retrieve under some time series models. For instance, with stationary AR processes, in one-step-forward forecasting, $R_{T,cond}'$ is the ratio between the variance of the innovation term and the marginal variance of the process.

\section{Empirical application}\label{sec:4}

Now we illustrate our proposed methodology on the SAL dataset presented in the introduction. We ran all our analyses in the statistical computing environment \texttt{R} \cite{Rstats}. We used the package \texttt{gstat} \cite{gstat} to compute a spatio-temporal variogram and then to perform variogram fitting. The first operation was the most compute-intensive and memory-consuming, and it was done with the aid of a virtual machine on Microsoft Azure\footnote{Microsoft Azure is a cloud computing service, see \url{www.azure.microsoft.com}.}. All the other analyses were instead carried out using typical laptop computers.

\subsection{Explorative analysis}

\begin{figure}
	\centering
	\caption{\label{fig:temptrends1} Univariate time series of temperature per sensor, training set (October through December). Peaks and troughs are thresholded and highlighted with dots.}
	\includegraphics[width=0.8\linewidth]{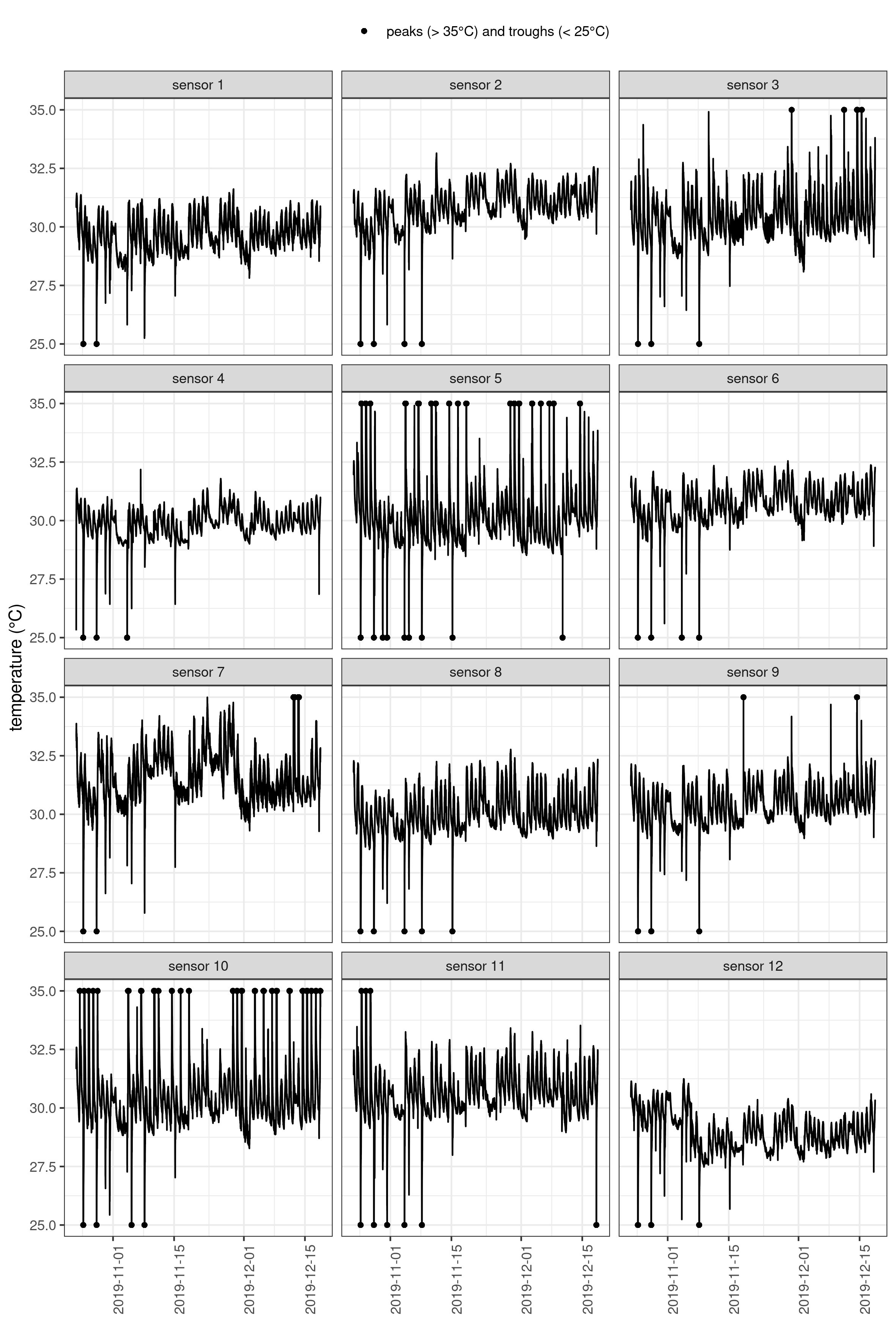}
\end{figure}

\begin{figure}
	\centering
	\caption{\label{fig:temptrends2} Sample quantiles of temperature, aggregation according to the weekday, training set (October through December).}
	\includegraphics[width=0.8\linewidth]{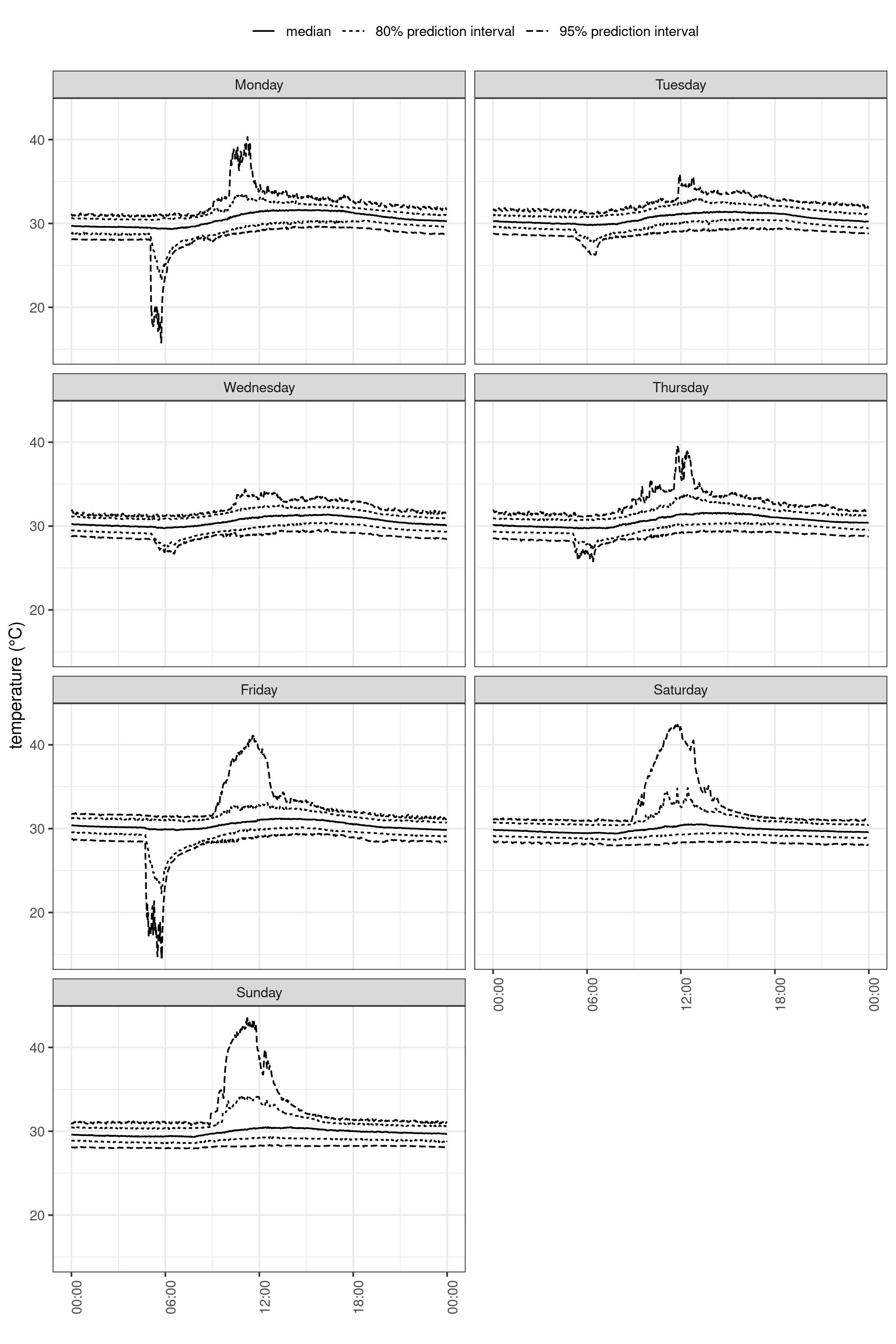}
\end{figure}

The SAL dataset collects temperature sensor readings from an office room and has been briefly described in the introductory section. The goal is to develop a spatio-temporal prediction rule for temperature based on these data. Some candidate spatial models are estimated on a training set, and the best one is selected based on a test set. The full dataset comprises 19 weeks of data and it is partitioned accordingly, with the leading eight weeks of data for training and the trailing 11 weeks as the test set. This choice is challenging for our method, as it is more exposed to shifts in the regime, but a longer training phase might not be reasonable for some applications, where a monitoring system shall be calibrated in short amounts of time.

Data were missing less than 1\% of times, resulting from miscommunication faults unrelated to the data and thus statistically random. Kriging can handle missings at random, but we used a simpler imputation method called \textit{last observation carried forward} (LOCF), which uses the last valid reading to impute missings \cite{Zhou_2018}. In fact, we require gridded data and the LOCF approach solves the problem rapidly and efficiently, so we can focus on other aspects of the problem.

Figure \ref{fig:temptrends1} presents the training set. The sensors are numbered from 1 to 12 as in Figure \ref{fig:room}. Troughs are concentrated in the mornings, as windows are opened, and cold air flows in the room during routine cleaning. Peaks are concentrated around noon, as direct sunlight overheats the sensors facing south, the ones numbered 5, 10 and 11. Weekly trends are highlighted in Figure \ref{fig:temptrends2}, with temperature median and other percentiles reported throughout weekdays. Troughs seem to occur mostly on Mondays and Fridays, so on the first and last workdays in the week. Peaks instead concentrate on Fridays and weekends.

Similar conditions between subsequent days motivate using time-series models with seasonal components, the period being one day long. Similar events occurring on the same weekday motivate considering one more seasonal component, whose period should be one week long. This kind of seasonal modeling is not considered in variogram fitting since it would be cumbersome to evaluate the ACF of a general model as per Equation \eqref{model:ar}. Instead, considering seasonal models is feasible with our inferential and predictive approach.

Spatial and temporal dependence seem able to explain most of the variability of temperature data. One may consider adding covariates into the model, in particular the physical variables mentioned in the introduction. Some preliminary analysis show that their explanatory power is limited, though, so they will not be considered further.

\subsection{Model estimates and comparison with variogram fitting}

We compare our proposed composite likelihood estimation approach with least squares variogram fitting (VF), which is a standard in the estimation of spatio-temporal correlation structures \cite{Cressie_1985}. VF requires to compute the empirical spatio-temporal variogram. The temporal covariance is feasibly tapered by considering only data pairs with time lag less than some custom threshold $T_{max}$. A variogram model is then fitted to the empirical variogram in a non-linear least squares fashion. The empirical variogram is computed with cost $O(S^2\cdot T\cdot T_{max})$. With data of this size, VF seems very demanding for the implementation provided by the \texttt{R} package \texttt{gstat}.

In computing the empirical variogram, we used a virtual machine with four cores to emulate the computational power of the server. However, after running some pilot tests, we deemed it necessary to configure the virtual machine with 16 gigabytes of RAM, which would be overwhelming for typical Raspberry Pi boards. The variogram was computed using the eight weeks of training data, which took about 40 hours to complete with cores working in parallel. It is fair to notice that the comparison with \texttt{gstat} is somewhat uneven, because that library is general purpose and not optimized for separable models.

After calculating the empirical variogram, the subsequent step in VF is model fitting. The model was chosen to be separable, as discussed insofar. We consider four candidate models in the space domain, namely, exponential, Gaussian, power exponential and Matérn ACFs. As to the time domain, \texttt{gstat} supports the exponential ACF, which implies an AR(1) model with structural lag equal to the sampling interval, that is, $\Delta_1=10$ seconds in Equation \eqref{model:ar}. Fitting the variogram models took just a few seconds, as contrasted to computing the empirical variogram, but only boundary estimates were obtained for either the temporal correlation parameters or the spatial ones.

Unfortunately, these boundary results make comparisons between VF and our estimation approach not very informative. We then carried out maximum composite likelihood estimation of the same spatial and temporal models as with VF. The estimation of spatial parameters is based on the sample spatial correlation matrix, whose computational cost is $O(S^2\cdot T)$. Temporal parameters are estimated by iteratively transforming data and the number of iterations is generally stochastic, but fitting an AR(1) is simple because it involves evaluating the empirical ACF only once and at a specific lag. After computing the sample correlation matrix, the spatial composite likelihoods are maximized for each of the four candidate spatial models. As a remark, the temporal model was estimated once, independently of the four candidate spatial models, as they are not involved in the temporal composite likelihood.

\input{images/estsummary.tex}

\begin{figure}
	\centering
	\caption{\label{fig:acfs} Empirical auto-correlation (dots) and estimated ACFs for two spatial models (lines).}
	\includegraphics[width=0.9\linewidth]{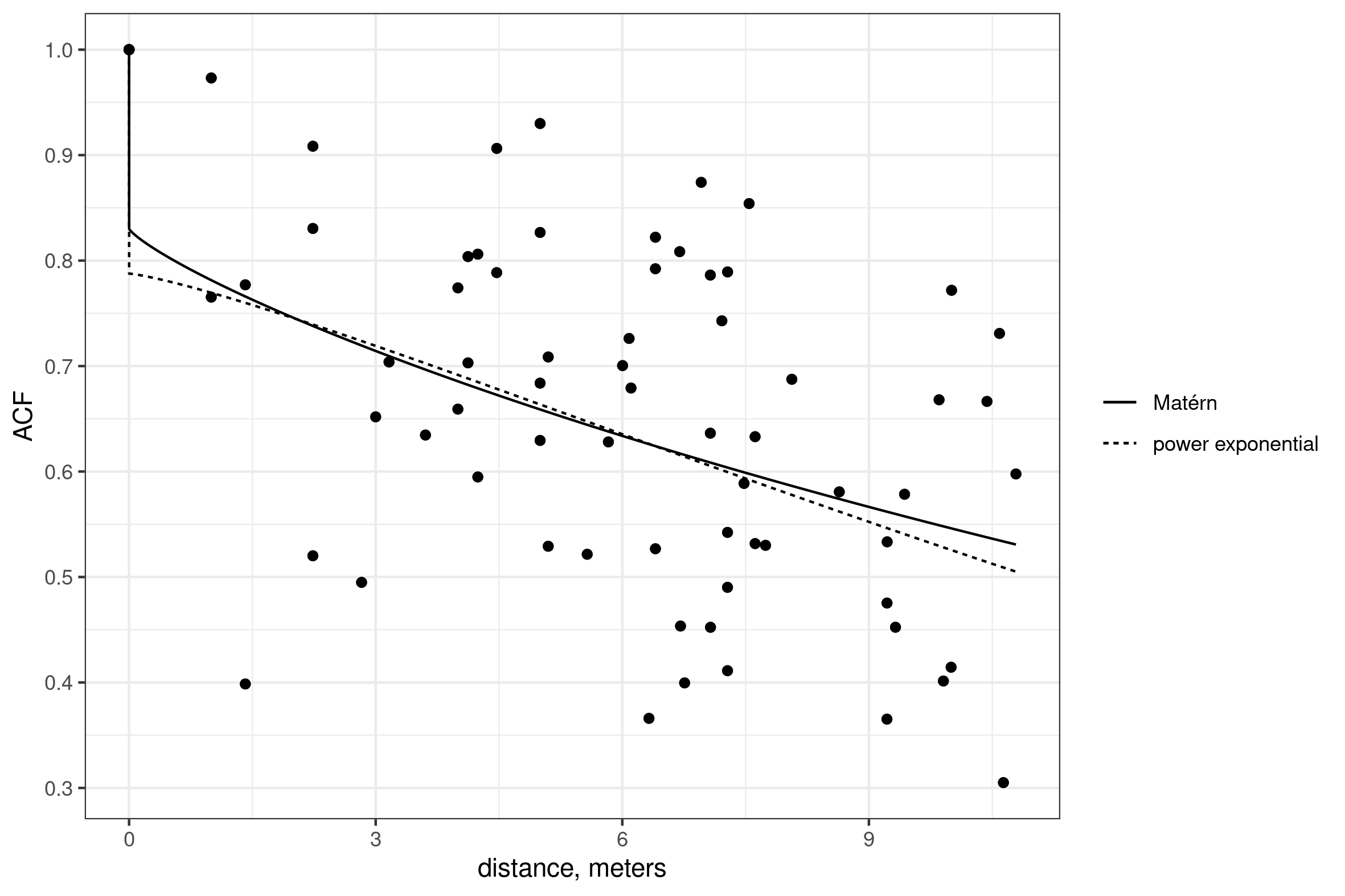}
\end{figure}

The estimate based on the composite likelihood for the AR coefficient is also boundary, i.e. close to unit. This result reflects the high sampling rate and the likely strong auto-correlation in the process. We then use longer structural lags, so that the model can be used even at lower sampling rates. In particular, we consider a custom AR model with lags $\Delta_1=10$ minutes, $\Delta_2=1$ day and $\Delta_3=1$ week in Equation \eqref{model:ar}. This model allows to assess multiplicative daily and weekly seasonal effects. It would be challenging to estimate with VF, as it relies on the ACF, which is hard to formulate with such a model. After estimating the custom AR and the four spatial models, we carry out inference based on parametric bootstrap by simulating and analyzing 1000 datasets for each joint spatio-temporal model. Some details on this procedure are given in Appendix \ref{app:bootstrap}. Performing the bootstrap took about 3 hours on a single laptop computer with our composite likelihood approach, despite the large sample size, while it would have been out of reach with VF. The model fit is summarized in Table \ref{tab:ests}. The implied spatial ACFs are reported in Figure \ref{fig:acfs} along with the empirical correlogram.

\subsection{Sensor network optimization}\label{sec:selection}

After estimating the temporal and spatial models, a practical concern is the selection of few operation sensors from the initial set, as twelve of them are too many for a room that is $127 m^2$ large. Under the proximity principle, some sensors could be dropped, and their location could be just virtually sensed since their data can be \textit{surrogated} \cite{Gramacy_2014} with predictions from the remaining sensors.

\begin{figure}
	\centering
	\caption{\label{fig:selection} Forward selection of sensors, distinct per metric, prediction error versus number of selected sensors. The sensor added at each step appears within a box. Run on test set (December through March).}
	\includegraphics[width=0.9\linewidth]{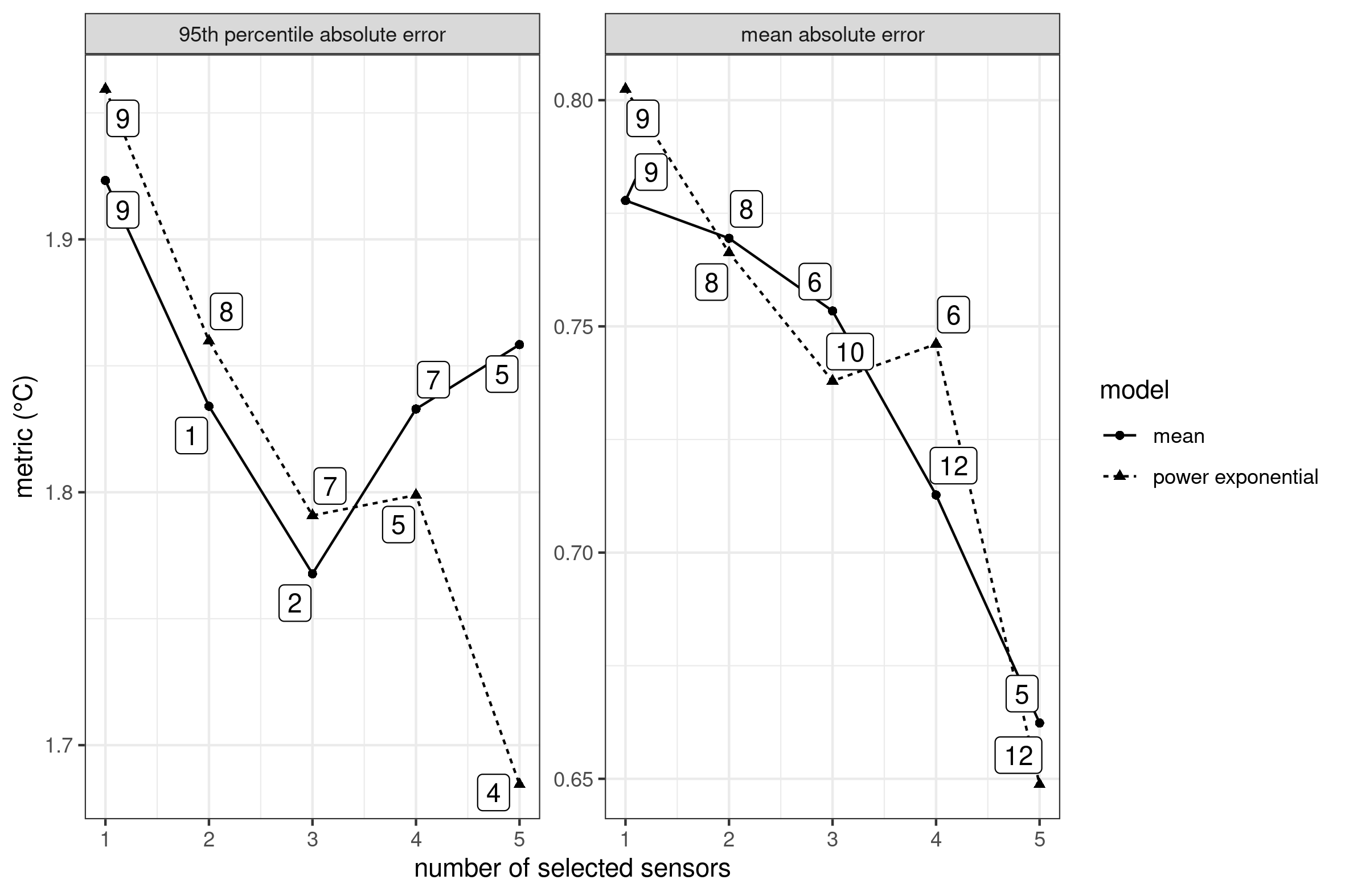}
\end{figure}

Different network configurations can be evaluated and compared according to a metric, which should reflect priorities and objectives of stakeholders. The 95th percentile of absolute prediction errors on all but active sensors \cite{Brunello2021} can be used for an approximate minimax decision. For comparison, we illustrate a sensor selection based on this criterion alongside one that uses a more classical mean absolute prediction error. The performance of spatial models and sensor configurations is evaluated and compared on the test set. For each sensor configuration, we interpolate data from selected locations to the unselected ones within time frames, as implied by separability. Prediction errors are then summarized according to the metric. We perform selection in a forward fashion, by starting with the best performer alone and then adding the sensor that led to the best improvement at each step. Adding sensors can worsen the performance because we are evaluating models on the test set. In Figure \ref{fig:selection}, a summary of the selection process is reported. The sensor added at each step appears within a box and is numbered as in Figure \ref{fig:room}. An alternative prediction is given by the simple mean, which assumes that a single latent temperature is ruling the whole room. The selection took no more than 10 minutes in total, so it would be easy to perform it multiple times \textit{ad interim} to check on the quality of predictions.

We show only the power exponential ACF. The result based on the Matérn function is very similar, and the exponential and Gaussian ones are slightly outperformed. The percentile performance seems in line with $k$-NN and IDW benchmarks \cite{Brunello2021}. Based on performances in Figure \ref{fig:selection}, the power exponential ACF may be preferred over the mean prediction because this choice seems more robust with respect to the metric. Moreover, the mean prediction yields some narrowly spaced sensor configurations under both metrics. 

\subsection{Behavior of the monitoring system}

\begin{figure}
	\caption{\label{fig:sampleperf} Examples of prediction via four sensors, based on power exponential model. Forecasts for February 17, 12:00, based on data from 10 minutes before. Circles are colored based on \textit{observed} temperatures, the floor based on \textit{predicted} temperatures, selected sensors only are numbered as in Figure \ref{fig:room}.}
	\begin{subfigure}{0.5\linewidth}
		\includegraphics[width=\linewidth]{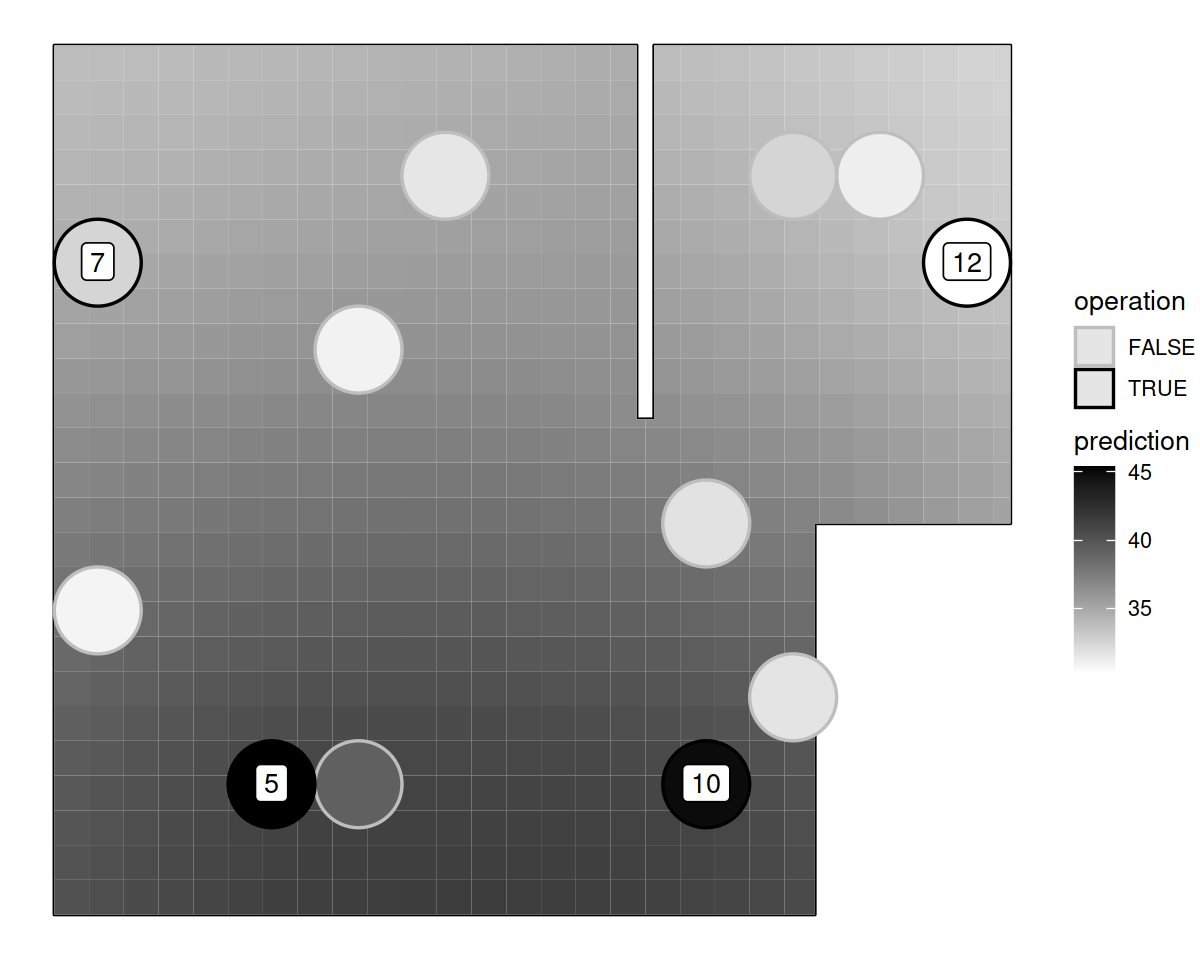}
		\caption{\label{fig:sampleperfanomalous} Anomalous sensors.}
	\end{subfigure}
	\begin{subfigure}{0.5\linewidth}
		\includegraphics[width=\linewidth]{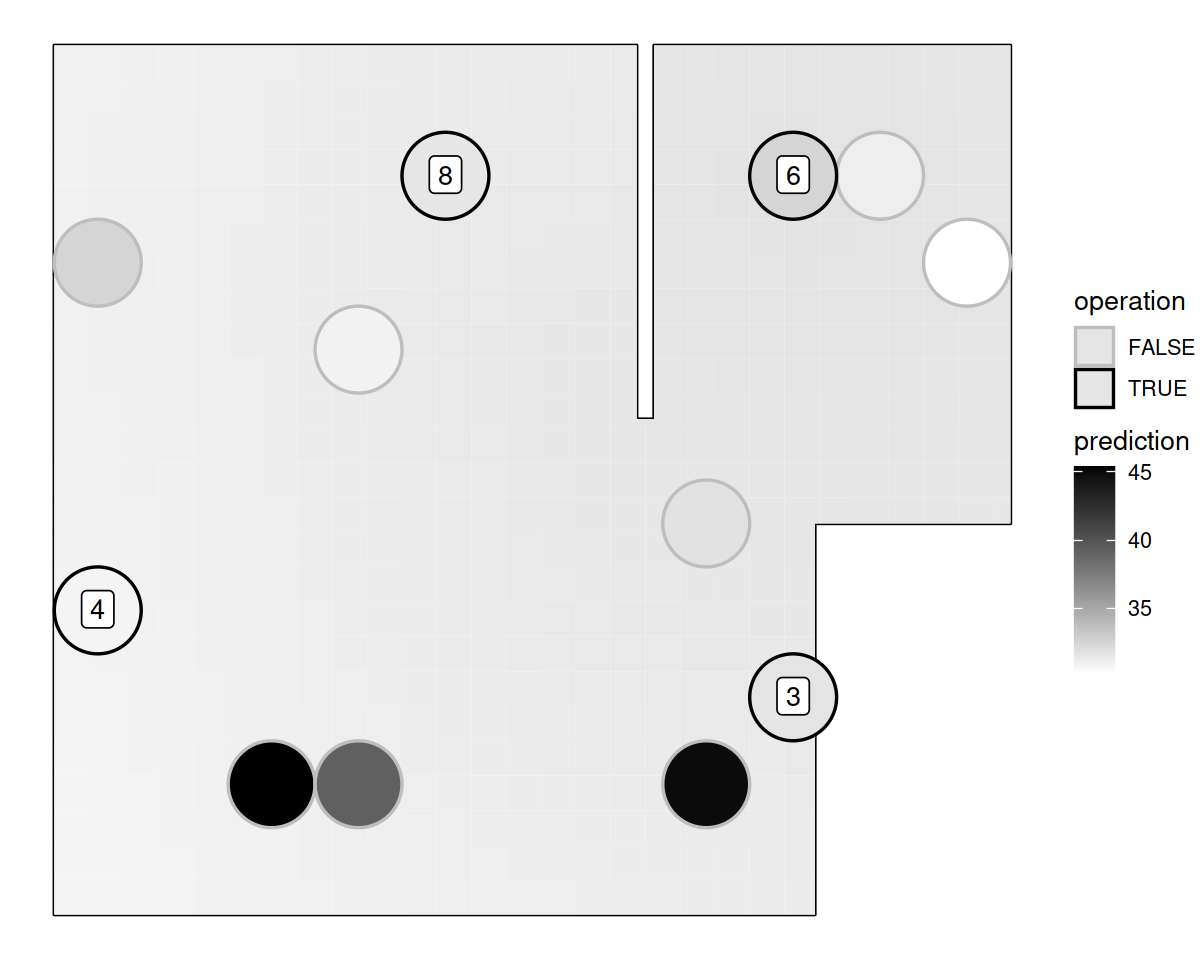}
		\caption{\label{fig:sampleperfregular} Regular sensors.}
	\end{subfigure}
	\begin{subfigure}{0.5\linewidth}
		\includegraphics[width=\linewidth]{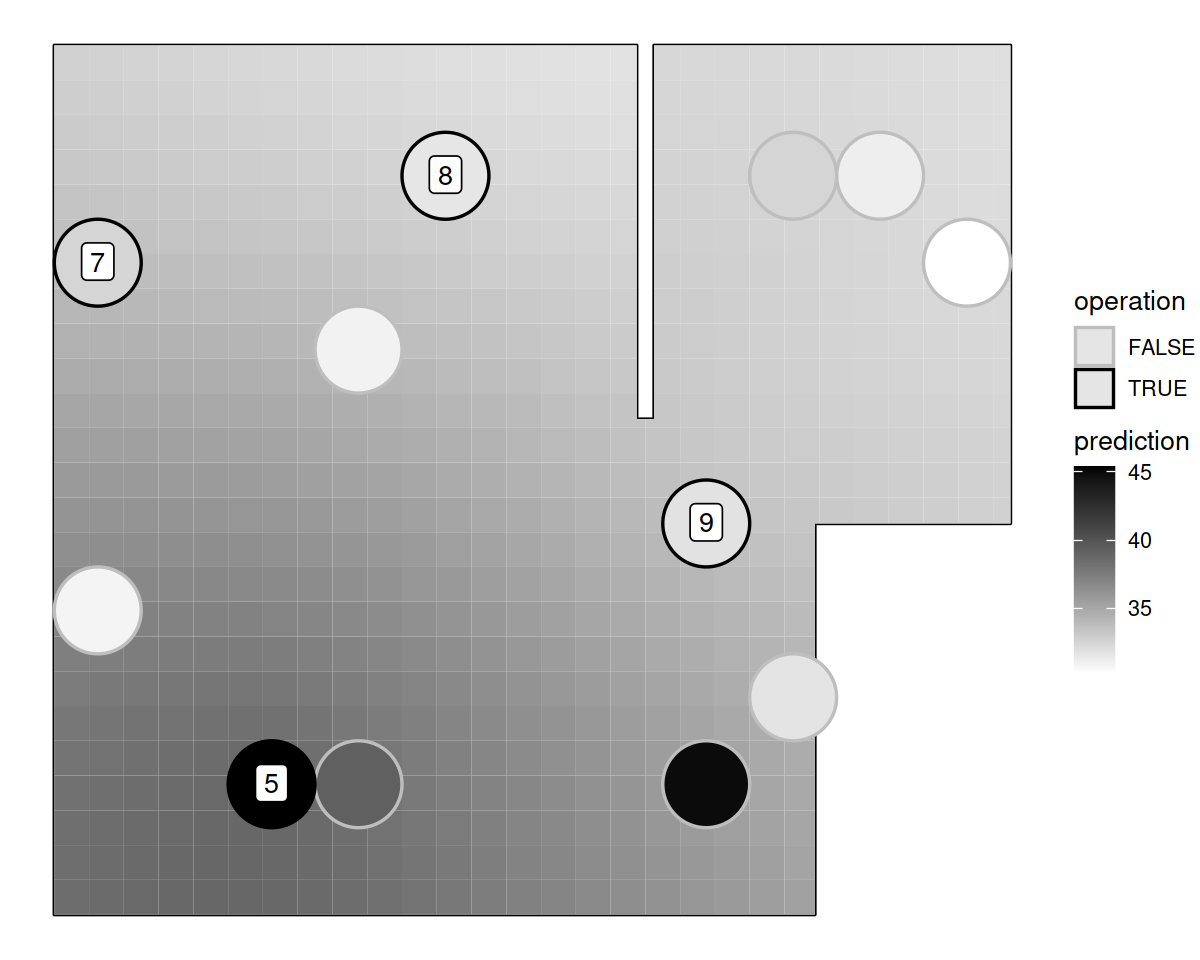}
		\caption{\label{fig:sampleperfpercen} Selection via 95th percentile absolute error.}
	\end{subfigure}
	\begin{subfigure}{0.5\linewidth}
		\includegraphics[width=\linewidth]{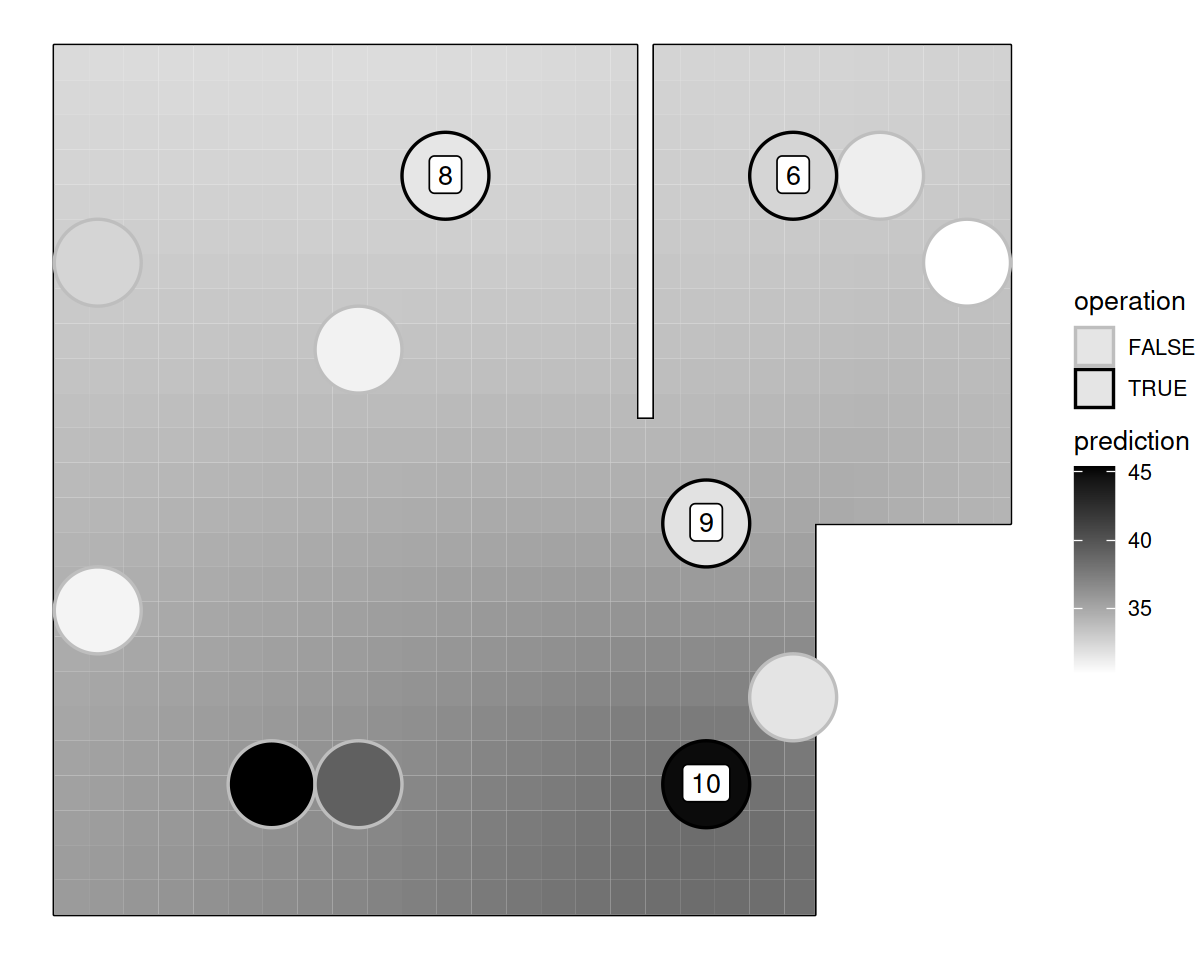}
		\caption{\label{fig:sampleperfmeanabs} Selection via mean absolute error.}
	\end{subfigure}
\end{figure}

By using different sensor configurations, we show few examples of actual situations along with the behavior of a possible monitoring system. The environment in focus has some peculiarities that somehow affect the performance of the system. As an illustration, we combine interpolation and forecasting by predicting temperatures 10 minutes forward for the whole floor plan, as in Figure \ref{fig:sampleperf}.

We anticipated that there would be differences between regular and anomalous sensors. Concerns relate to sensors facing direct sunlight around noon (numbered 5, 10) or close to other sources of anomalies (7, 12). If these were used to make predictions, the results would hardly reflect the normalcy ruling the interior of the room, see Figure \ref{fig:sampleperfanomalous}. On the contrary, a system based on more regular sensors only, like in Figure \ref{fig:sampleperfregular}, would yield more constant predictions that are likely exchangeable with mean prediction. The sensor selection illustrated in the previous section is aimed at selecting a solution between these two extremes.

The two metrics considered suggest similar solutions. Figure \ref{fig:sampleperfpercen} reports the prediction provided under the power exponential ACF by the four best sensors according to the percentile metric. Figure \ref{fig:sampleperfmeanabs} shows the selection based on the mean absolute error instead. Both configurations attempt to replicate the north-south gradient, which seems to require choosing between sensors 5 or 10. However, neither of these choices can surrogate sensors 11 and 3. Selecting sensors 5 and 10 would fail the whole interior of the room and, by converse, the sensors in the interior of the room cannot surrogate 5 and 10.

\section{Some extensions}\label{sec:5}

In the previous section, we presented an application of our compute-efficient approach to spatio-temporal kriging. Here we describe two possible extensions that look legitimate under our prediction-oriented view of kriging. For instance, it is possible to include some spatial interpolators or temporal forecasting methods that do not necessarily underlie any stationary ACF. It is also possible to use distinct temporal models for each sensor location to provide more specialized forecasting. Both these extensions share at least the advantage of further distributing calculus across the sensor network and simplify server-side computations.

\subsection{Non-stationary modeling}

The kriging approach relies on stationary ACF models, which offer a wide variety of possibilities, but some problems may be addressed only with non-stationary models. For instance, integrated AR models need no trend formulation. Another alternative is $k$-NN, which returns the sample average of response values from the $k$ sensors closest to the desired location.

This could be useful in cases like ours, where distinct sensors might have different equilibria. We used a moving average to accommodate for non-stationarity in the mean, which conciliates with stationarity in ACF, but one can use integrated AR and $k$-NN as an alternative. With high-frequency data, long-term stationarity may coexist with short-term non-stationarity, the latter resulting from locally linear trends, so it might be useful to consider a non-stationary model that copes with both aspects.

We considered an integrated AR model in our analysis but without obtaining any significant improvement. Indeed, the chosen AR model was already able to cope with non-stationarity due to both a moving average trend and a near-unit first AR coefficient, which implied integration \textit{de facto}.

\subsection{Sensor-specific temporal correlation parameters}\label{sec:sensspecmodel}

A limitation of separable ACFs is that they imply all locations having the same marginal dynamics. As an extension, the temporal correlation parameters can be sensor-specific: each sensor can estimate and update a distinct temporal model that is valid at least for its location and approximately also for a neighborhood.

Distinct temporal models can each be based on a different composite likelihood and use different portions of data, so they will not affect each other. In mean prediction, as per Equation \eqref{eq:krigtemp}, $\hat{z}$ can be replaced with a matrix, where each column is made up of the temporal forecasts based on a model with limited scope that works for just one location or a neighborhood. When interpolating these forecasts spatially, via Equation \eqref{eq:krigspat}, more weight is given to forecasts close to the needed locations. This implies that all temporal models are involved but to a varied extent, depending on the distance.

This extension with distinct temporal models per location adds flexibility to monitoring in at least two ways.
\begin{itemize}
	\item It adds flexibility in network management. Each sensor has to estimate and update its own temporal model, so this has not to be handled by the server, which thus must be in charge only with the spatial interpolation task.
	\item Statistical modeling becomes more flexible too. Prior to this, a single overall temporal model is formulated that has to fit all locations forcefully. Distinct temporal models may now be used to address subsets of locations, so they can cope with more local and specific dynamics.
\end{itemize}

Anomalous sensors may be more effectively dealt with by allowing them to make predictions based on a more specific model targeted to them only. The office room in our example is too small to allow for a variety of temporal models. Larger environments will likely be more heterogeneous and will thus need many local models to provide better forecasts. Indeed, since spatio-temporal prediction is made up of both interpolation and forecasting, the quality of the latter is a necessary ingredient to joint predictive performance.

\section{Discussion}

We have proposed a separable kriging approach that allows to analyze large datasets by exploiting some overlooked aspects of separability. Even high-frequency data can be processed in a reasonable amount of time using a maximum composite likelihood estimator and optimized calculus in prediction. Separability allows to distribute calculus across the sensor network by delegating as many operations as possible to the components that gather the relevant data.

To our knowledge, the use of marginal composite likelihood is novel to sensor data analysis. Its most appealing aspect is that the spatial and temporal models under separability can be estimated in parallel without affecting each other. The spatial model must be estimated in a centralized way, but the temporal model may be addressed in a decentralized way by allowing sensors to estimate a temporal model valid for their location or neighborhood, as described in Section \ref{sec:sensspecmodel}. This idea relates to stratified variograms \cite{Courault_1999,Bivand_2013}, but it has even more in common with the estimation of a single variogram with data pairs sharing some identical conditions \cite{Monestiez_2001}.

The predictive part of our approach was already common in climate and weather sciences, though in a modeling-unaware fashion, and confined mostly to spatial interpolation \cite{Holdaway_1996}. Instead, we provide a formal motivation for this way of computing predictions based on separability. We found a related simplification in jointly spatio-temporal prediction, which we guess can be easily extended to separability in more than two domains via Tucker products instead of Kronecker ones. For instance, covariates could be included in kriging via fully-factored modeling \cite{Mardia1993}, but feature engineering seemed necessary in our case \cite{Brunello2021}, which is not typical in kriging. For the sake of completeness, alternative modeling strategies include additive covariances \cite{Ma_2019}, process convolution \cite{Higdon_2002} and linear mixed models \cite{Dumelle_2021}, for which simplifications might be different where possible.

In developing our proposal, we require data to be gridded, which means that all sensors provide simultaneous readings. However, this requirement can be weakened, since data can be at least projected onto a grid \cite{Paciorek_2007}.

Kriging computation is hugely simplified by assuming separability and by choosing a suitable spatial or (especially) temporal model. Both estimation and prediction can bypass the evaluation and inversion of large correlation matrices by treating the data as univariate time series or cross-sections and thus splitting a generally complicated calculation into simpler operations. For instance, AR models may have an intractable ACF, but they can be estimated easily by minimizing a conditional sum of squares, and their forecasts based on Equation \eqref{model:ar} are simple to calculate as well. Our approach allows to blend together and leverage on known strengths of time series analysis and spatial statistics, without the need to outline a joint spatio-temporal framework from scratch.

\section{Acknowledgements}


This project was performed within the COMET Centre ASSIC Austrian Smart Systems Integration Research Center, which is funded by BMK, BMDW, and the Austrian Provinces of Carinthia and Styria, within the framework of COMET -- Competence Centers for Excellent Technologies. The COMET programme is run by FFG. The research of Michele Lambardi di San Miniato was supported by the European Social Fund (Investimenti in favore della crescita e dell'occupazione, Programma Operativo del Friuli Venezia Giulia 2014/2020) - Programma specifico 89/2019 - Sostegno alla realizzazione di dottorati e assegni di ricerca, operazione PS 89/2019 ASSEGNI DI RICERCA - UNIUD (FP1956292002, canale di finanziamento 1420\_SRDAR8919).

No conflict of interest has been detected.

The dataset used in the application example is publicly available on GitHub, as submitted by the authors of the first paper addressing it \cite{Brunello2021}, at
\begin{center}
	\url{https://github.com/dslab-uniud/virtual-sensing} .
\end{center}

\appendix

\section{Proofs}\label{app:krig}

\subsection{Kriging mean formula}\label{app:krigmean}

Predictions can be computed in a vectorized form, as follows, after Equation \eqref{eq:kriging}.
$$\vec(\hat{y}'-\mu') = \rho R^{-1} \mathrm{vec}(y-\mu) \,.$$
By using definitions in Equations \eqref{eq:kron} and \eqref{eq:otherkronecker}, it follows that
$$\vec(\hat{y}'-\mu') = (\rho_S \otimes \rho_T)  (R_S\otimes R_T)^{-1}  \mathrm{vec}(y-\mu) \,.$$
Next, we use the inversion behavior of the Kronecker product.
$$\vec(\hat{y}'-\mu') = (\rho_S\otimes \rho_T)  (R_S^{-1} \otimes R_T^{-1})  \mathrm{vec}(y-\mu) \,.$$
Then, it comes in handy to use the mixed-product property of the Kronecker product.
$$\vec(\hat{y}'-\mu') = \left\{(\rho_S R_S^{-1}) \otimes (\rho_T  R_T^{-1}) \right\} \mathrm{vec}(y-\mu) \,.$$
At this point, the regression coefficients of Equation \eqref{eq:regcoef} can be recognized.
$$\vec(\hat{y}'-\mu') =  (\beta_S \otimes \beta_T)  \mathrm{vec}(y-\mu) \,.$$
Lastly, we use Roth's column lemma \cite{Hartwig_1975}.
$$\vec(\hat{y}'-\mu') =  \mathrm{vec}\left\{\beta_T (y-\mu)  \beta_S^\top\right\} \,.$$
Then, the $\vec$ operator can be dropped and the matrix $\hat{y}'$ is obtained.

\subsection{Kriging variance formula}\label{app:krigvar}

Using Equation \eqref{eq:kriging} as a starting point, it holds
$$\mathrm{diag}(R_{cond}') = \mathrm{diag}(R') - \mathrm{diag}(\rho R^{-1} \rho^\top) \,.$$
Similarly to the proof in Appendix \ref{app:krigmean}, we exploit again the inversion behavior and the mixed-product property of the Kronecker product. We also use Equations \eqref{eq:kron} and \eqref{eq:otherkronecker} to obtain
$$\mathrm{diag}(R_{cond}') = \mathrm{diag}(R_S' \otimes R_T') - \mathrm{diag}\left\{(\rho_S R_S^{-1} \rho_S^\top) \otimes (\rho_T R_T^{-1} \rho_T^\top) \right\} \,.$$
After Equation \eqref{eq:condcors}, it follows that
$$\mathrm{diag}(R_{cond}') = \mathrm{diag}(R_S' \otimes R_T') - \mathrm{diag}\left\{(R_S' - R_{S,cond}') \otimes (R_T' - R_{T,cond}') \right\} \,.$$
One can use the self-evident property $\mathrm{diag}(A \otimes B) = \mathrm{diag}(A) \otimes \mathrm{diag}(B)$ for $A$ and $B$ square matrices, which implies
$$\mathrm{diag}(R_{cond}') = \mathrm{diag}(R_S') \otimes \mathrm{diag}(R_T') - \mathrm{diag}(R_S' - R_{S,cond}') \otimes \mathrm{diag}(R_T' - R_{T,cond}') \,.$$

Now, the components of $\mathrm{diag}(R_{cond}')$ can be partitioned into vectors with the same length as $\mathrm{diag}(R_T')$. Such vectors can be the columns of the matrix $V$, which is thus defined as in our claims.

\section{Bootstrap}\label{app:bootstrap}

Parametric bootstrap \cite{Davison_1997} under separable kriging is particularly convenient because the implied model is easy to simulate, and its parameters are simple to estimate with the approach proposed in this paper.

The temporal and spatial model together identify the full model, under which one can simulate artificial datasets. In particular, separability allows to simulate a dataset $Y$ as
$$Y \sim \mu + \sigma \cdot R_T^{1/2} \cdot \epsilon \cdot R_S^{1/2} \,,$$
where $\epsilon$ is a Gaussian white noise structured into a $T \times S$ matrix, and $R_T^{1/2}$ and $R_S^{1/2}$ are the matrix square roots of the matrices $R_T$ and $R_S$, respectively.

Assuming $T \gg S$, $R_S^{1/2}$ may be tractable, while $R_T^{1/2}$ will hardly be so. The operator $R_T^{1/2}$ just makes $R_T^{1/2} \cdot \epsilon$ a matrix with independent columns that share the same correlation structure $R_T$. So, as an alternative to directly evaluating $R_T^{1/2}$, one can generate each column of $\sigma \cdot R_T^{1/2} \cdot \epsilon$ according to the temporal model. AR(p) processes can be simulated efficiently according to the factorized MA($\infty$) form \cite{Box}. The first observations should be initialized according to the stationary distribution of the process, but with complicated models one may instead provide an arbitrary initialization and then simulate additional observations as a burn-in. We adopted this latter strategy. Actually, in simulating eight weeks of data, we needed to simulate 32 more leading weeks of data as a burn-in.

\end{document}

%% file: images/estsummary.tex
\begin{table}

\caption{\label{tab:ests}Point estimates and standard errors for spatial and temporal correlation parameters.}
\vspace{1em} \centering
\begin{tabular}[t]{l|l|r|r}

spatial model & parameter & est. & std. err. ($\times 10^3$)\\
\hline
 exponential & $\phi_1$ & 0.977 & 0.206\\

 & $\phi_2$ & 0.078 & 0.950\\

 & $\phi_3$ & 0.047 & 0.990\\

 & nugget & 0.191 & 2.307\\

 & range & 24.692 & 446.329\\

\hline
 Gaussian & $\phi_1$ & 0.977 & 0.203\\

 & $\phi_2$ & 0.078 & 0.936\\

 & $\phi_3$ & 0.047 & 0.975\\

 & nugget & 0.250 & 2.470\\

 & range & 15.011 & 140.078\\

\hline
 Matérn & $\phi_1$ & 0.977 & 0.206\\

 & $\phi_2$ & 0.078 & 0.950\\

 & $\phi_3$ & 0.047 & 0.991\\

 & nugget & 0.187 & 4.805\\

 & range & 25.993 & 1547.047\\

 & smoothness & 0.479 & 22.544\\
\hline
 power exponential & $\phi_1$ & 0.977 & 0.205\\

 & $\phi_2$ & 0.078 & 0.946\\

 & $\phi_3$ & 0.047 & 0.986\\

 & nugget & 0.217 & 3.057\\

 & range & 19.883 & 424.822\\

 & smoothness & 1.312 & 28.654\\

\end{tabular}
\end{table}